\documentclass[sigconf]{acmart}
\usepackage{graphicx}
\usepackage{float}
\usepackage{amsthm,amsmath}

\newtheorem{thm}{Theorem}

\usepackage{graphicx}       
\usepackage{subfig}
\usepackage{hyperref}
\usepackage{tabularx}
\usepackage{multirow}  

\usepackage{amsthm,amsmath}
\usepackage{mathrsfs}
\usepackage{bm}
\usepackage{enumitem}

\usepackage{makecell} 

\newcommand{\ie}{\emph{i.e., }}
\newcommand{\eg}{\emph{e.g., }}

\newcommand{\cf}{\emph{cf. }}

\AtBeginDocument{%
  }

\copyrightyear{2025}
\acmYear{2025}
\setcopyright{acmlicensed}
\acmConference[WWW Companion '25]{Companion Proceedings of the ACM Web Conference 2025}{April 28-May 2, 2025}{Sydney, NSW, Australia}
\acmBooktitle{Companion Proceedings of the ACM Web Conference 2025 (WWW Companion '25), April 28-May 2, 2025, Sydney, NSW, Australia}
\acmDOI{10.1145/3701716.3715548}
\acmISBN{979-8-4007-1331-6/2025/04}
\begin{document}

\title{Debias Can Be Unreliable: Mitigating Bias in Evaluating Debiasing Recommendation}


\author{Chengbing Wang} 
\email{wwq197297@mail.ustc.edu.cn}
\affiliation{%
  \institution{University of Science and Technology of China}
  \city{Hefei}
  \country{China}
}
\author{Wentao Shi} 
\email{shiwentao123@mail.ustc.edu.cn}
\affiliation{%
  \institution{University of Science and Technology of China}
  \city{Hefei}
  \country{China}
}

\author{Jizhi Zhang} 
\email{cdzhangjizhi@mail.ustc.edu.cn}
\affiliation{%
  \institution{University of Science and Technology of China}
  \city{Hefei}
  \country{China}
}

\author{Wenjie Wang} 
\email{wenjiewang96@gmail.com}
\affiliation{%
  \institution{National University of Singapore}
  \city{Singapore}
  \country{Singapore}
}
\authornote{Corresponding author}

\author{Hang Pan} 
\email{hungpaan@mail.ustc.edu.cn}
\affiliation{%
  \institution{University of Science and Technology of China}
  \city{Hefei}
  \country{China}
}

\author{Fuli Feng} 
\email{fulifeng93@gmail.com}
\affiliation{%
  \institution{University of Science and Technology of China}
  \city{Hefei}
  \country{China}
}
\authornotemark[1]

\renewcommand{\shortauthors}{Chengbing Wang et al.}
\begin{abstract}
Recent work has improved recommendation models remarkably by equipping them with debiasing methods. 
Due to the unavailability of fully-exposed datasets, most existing approaches resort to randomly-exposed datasets as a proxy for evaluating debiased models, employing traditional evaluation scheme to represent the recommendation performance. However, in this study, we reveal that traditional evaluation scheme is not suitable for randomly-exposed datasets, leading to inconsistency between the Recall performance obtained using randomly-exposed datasets and that obtained using fully-exposed datasets. To bridge the gap, we propose the \underline{U}nbiased \underline{R}ecall \underline{E}valuation (URE) scheme, which adjusts the utilization of randomly-exposed datasets to unbiasedly estimate the true Recall performance on fully-exposed datasets. We provide theoretical evidence to demonstrate the rationality of URE and perform extensive experiments on real-world datasets to validate its soundness. Our code and datasets are available at \url{https://anonymous.4open.science/r/URE-6B64/}.

\end{abstract}





\maketitle

\section{Introduction}

Recommendation models suffer from various biases
leading to deviations from true user preferences even harmful impacts to Web services~\cite{DBLP:journals/tois/0007D0F0023,pop_1,DBLP:conf/icml/SchnabelSSCJ16,DBLP:conf/icml/LiXZ0023,DBLP:conf/ijcai/Pan0FSW023}.
Undoubtedly, user's true preferences can be accurately measured by the fully-exposed dataset which includes users' feedback (\eg ratings) across all items. 
The gold standard for evaluating debiasing techniques is thus utilizing such a fully-exposed dataset to assess the performance of recommendations~\cite{DBLP:conf/cikm/GaoLLCLJ0MC22}.
However, due to the unavailability of fully-exposed datasets in practice~\cite{DBLP:conf/cikm/GaoLZCLL0022}, randomly-exposed datasets, which contain users' feedback on some randomly selected items, are widely employed for debiasing evaluation~\cite{DBLP:conf/icml/SchnabelSSCJ16,DBLP:conf/sigir/LiuCDHP020,DBLP:conf/www/RosenfeldMY17}. Notably, the amount of feedback in randomly-exposed datasets is extremely limited, because randomly exposing items to collect user feedback is highly expensive, and the collection process can severely damage a platform's revenue. 


Previous work calculate widely-used Recall metric on randomly-exposed datasets in the same way as on fully-exposed datasets, which we refer to as the traditional evaluation scheme in subsequent descriptions. Nevertheless, we claim that traditional evaluation scheme is not suitable for computing Recall metric on randomly-exposed datasets when evaluating debiasing recommendation. This is due to the significant disparities between the Recall obtained from the randomly-exposed dataset and those derived from the fully-exposed dataset. To comprehensively investigate the underlying factors contributing to the inconsistency, we undertake a theoretical analysis (\cf Theorem~\ref{theorem_K}) of the evaluation performed on a randomly-exposed dataset. Our analysis demonstrates that the Recall@$\overline{K}$ on the randomly-exposed dataset (\eg Recall@$\overline{5}$) has a weak correlation with the Recall@$K$ using a small value of $K$ on the fully-exposed dataset (\eg Recall@50). 
Although the correlation between Recall@$\overline{K}$ and Recall@$K$ could become stronger as the value of $K$ increases, the performance cutting at a small value of $K$ is more important in practice~\cite{DBLP:journals/tois/0007D0F0023}. 
In addition, we conduct extensive experiments on real-world datasets to validate such correlation (\cf Figure~\ref{fig:ranking}). Thus, the conclusion derived from the traditional evaluation scheme on the randomly-exposed dataset such as Recall@$\overline{5}$ might be unconvincing as it is inadequate to conclude whether previous debiasing methods have effectively mitigated biases.

Towards this goal, we propose the \underline{U}nbiased \underline{R}ecall \underline{E}valuation (URE) scheme. We set the target as estimating the Recall@$K$ on fully-exposed data, \ie the ratio of items with positive feedback ranked before position $K$. 
Considering that a randomly-exposed dataset only includes user feedback on randomly selected items, we can treat it as a representative subset of the fully-exposed dataset, wherein both positive and negative feedback are missing at random. In URE, we sort the model's prediction scores of all candidate items and utilize the score of $(K+1)$-th item as the threshold to calculate the positive ratio on the randomly-exposed dataset, where the positive ratio is the proportion of items with positive feedback in the randomly-exposed dataset that rank before the threshold.
Hence averaging such a positive ratio across all users can be used to unbiasedly estimate Recall@$K$ on a fully-exposed dataset. This is due to the models’ prediction information of all candidate items, which is not considered by traditional evaluation scheme. 
Through theoretical analysis and extensive experiments, we validate that URE can unbiasedly estimate the Recall@$K$ on fully-exposed data with a desired value of $K$. 

The main contributions are summarized as follows:
\begin{itemize}[leftmargin=*]

\item We reveal the inconsistency between the widely adopted traditional evaluation scheme on the randomly-exposed dataset and fully-exposed dataset (the gold standard evaluation), questioning empirical conclusions on existing debiasing methods in previous studies.
    

\item We propose the URE, a novel scheme to unbiasedly estimate the target Recall@$K$ performance with randomly-exposed data. 

\item We conduct extensive theoretical and empirical analysis, validating the effectiveness of the URE scheme. 

\end{itemize}

\section{Preliminary}

Given a fully-exposed dataset (denoted as $D_{full}$), we can determine the true performance of debiasing models by calculating the Recall formula, denoted as denoted as Recall@$K$:
\begin{equation}\label{eq3}
\begin{split}
    \text{Recall}@K = \frac{|{k\in Rank_{u}+: k \leq K}|}{|Rank_{u}+|},
    K \in [1,2,...,L],
\end{split}
\end{equation}
where the $Rank_{u}+$ compiles the positions of all items with positive feedback within the personalized ranking list recommended by the models.
However, The Recall@$K$ is not supported due to the unavailability of $D_{full}$ (The only existing $D_{full}$ for research is KuaiRec~\cite{DBLP:conf/cikm/GaoLZCLL0022}). In this case, researchers employ the same formula to compute Recall on a randomly-exposed dataset (denoted as $D_{rand}$). The resulting Recall is referred to as Recall@$\overline{K}$, where $\overline{K} \in [1,2,...,l], l<<L$. This evaluation approach for debiasing models is termed the "traditional evaluation scheme on $D_{rand}$".

\section{\mbox{Relation of $\text{Recall@}K$ and $\text{Recall@}\overline{K}$}}
\label{section:analysis 3}
In this section, we conduct theoretical analysis and empirical validation to discover why the traditional evaluation scheme on $D_{rand}$ is unreliable by inspecting the relationship between Recall@$K$ on $D_{full}$ and Recall@$\overline{K}$ on $D_{rand}$.

\subsection{Theoretical Guarantee}
In this subsection, we give the theoretical 
derivation of the relationship between Recall@$K$ on $D_{full}$ and Recall@$\overline{K}$ on $D_{rand}$.


\begin{thm}
Assuming that we have $N^{+}$ positive samples and $N^{-}$ negative samples on $D_{full}$, with a total sample size of $N=N^{+}+N^{-}$, our $D_{rand}$ samples $\overline{N}$ samples from $D_{full}$. We denote the set of all combinations of $D_{rand}$
of size $\overline{N}$ as $\mathcal{S}_{\overline{N}}$ and the set of all permutaions of $D_{full}$ as $\mathcal{R}$. Then, when $K$ on $D_{full}$ and $\overline{K}$ on $D_{rand}$ satisfy $\overline{K}=\frac{\overline{N}}{N}\cdot K$, we have:
\begin{equation}
    E_{\mathcal{R}}\left[ E_{\mathcal{S}_{\overline{N}}}[\text{Recall@}\overline{K}] - \text{Recall@}K\right] = 0,
    \label{eq:relation_Recall_K}
\end{equation}
\label{theorem_K}
where $E$ denotes the Expectation funtion.
\end{thm}

\label{proof_theorem_K}
\begin{proof}
We first calculate the $E_{\mathcal{R}}[\text{Recall@}K]$. Since we consider all possible ranking, the possibility of the x positive samples contained in the top K is $\frac{C_{N^+}^{x}C_{N - N^{+}}^{K-x}}{C_{N}^{K}}$ where $C^{n}_{N}:=\frac{N!}{(N-n)!n!}$ is the number of combinations obtained by randomly sampling $n$ samples from $N$. Thus, we have:
\begin{small}
\begin{align}
\label{eq:hyper_geo}
    & E_{\mathcal{R}}\left[\text{Recall}@K \right] = \sum_{x=1}^{min(K, N^{+})}\frac{x}{N^{+}}\frac{C^{x}_{N^{+}}C_{N-N^+}^{K-x}}{C_N^K} \\
    & = \frac{K}{N} \left[\sum_{x=1}^{min(K,N^+)}\left[C_{N^+-1}^{x-1} C_{(N-1) - (N^+-1)}^{(K-1) - (x-1)}/C^{K-1}_{N-1} \right] \right] \\
    & = \frac{K}{N}\left[\ C_{N-1}^{K-1}/ C_{N-1}^{K-1} \right] = \frac{K}{N}.
\end{align}
\end{small}

As for the $E_{\mathcal{R}}\left[ E_{\mathcal{S}_{\overline{N}}}[\text{Recall@}\overline{K}]\right]$, we have:
\label{eq:orig}
\begin{small}
\begin{align}
    &E_{\mathcal{R}}\left[ E_{\mathcal{S}_{\overline{N}}}[\text{Recall@}\overline{K}]\right] =  E_{{N^+}^{\prime}}\left[E_{\mathcal{R}, \mathcal{S}_{\overline{N}}}[\text{Recall@}\overline{K}|{N^+}^{\prime}]  \right].
\end{align}
\end{small}
Since we consider all possible permutations, $E_{\mathcal{R}, \mathcal{S}_{\overline{N}}}\left[\text{Recall@}\overline{K}|{N^+}^{\prime}\right]$ can be transformed into the scenario of randomly selecting $\overline{K}$ samples from $\overline{N}$ samples ($N^+$ positive samples among these $N$ samples) and determining how many of the selected $\overline{K}$ samples are positive. The problem is similar to Eq~\ref{eq:hyper_geo}. Thus we have:
\begin{small}
\begin{align}
    & E_{{N^+}^{\prime}}\left[E_{\mathcal{R}, \mathcal{S}_{\overline{N}}}[\text{Recall@}\overline{K}|{N^+}^{\prime}]  \right]\\
    &=E_{{N^+}^{\prime}}\left[\sum_{x=1}^{min(\overline{K},{N^+}^{\prime}) }\frac{x}{{N^+}^{\prime}}\frac{C^{x}_{{N^+}^{\prime}} C^{\overline{K} - x}_{\overline{N} - {N^+}^{\prime}}}{C_{\overline{N}}^{\overline{K}}}\right] \\
    & = E_{{N^+}^{\prime}}\left[\frac{\overline{K}}{\overline{N}}\frac{\sum_{x=1}^{min(\overline{K},{N^+}^{\prime})} C^{x - 1}_{{N^+}^{\prime} - 1} C^{(\overline{K} - 1) - (x - 1)}_{\overline{N} - {N^+}^{\prime}}}{C_{\overline{N}-1}^{\overline{K} - 1}}\right] \\
    & = E_{{N^+}^{\prime}}\left[ \frac{\overline{K}}{\overline{N}}\right] 
    = \frac{\overline{K}}{\overline{N}} = \frac{K}{N} = E_{\mathcal{R}}[\text{Recall@}K].
\end{align}
\end{small}

\end{proof}

From Theorem \ref{theorem_K}, we can observe that the Recall@$\overline{K}$ on $D_{rand}$ will exhibit a strong correlation with the Recall@$K$ on $D_{full}$ if the value of $K$ is around $\frac{{N}}{\overline{N}}\cdot \overline{K}$. 
It should be noted that $\overline{N}$ is much smaller than $N$ since the $D_{rand}$ is a sparsely sampled subset of $D_{full}$. 
Therefore, given a specific value of $\overline{K}$, the Recall@$\overline{K}$ on $D_{rand}$ only exhibits a strong correlation with the Recall@$K$ with much larger $K$ values on $D_{full}$.
However, given that user typically browse only a limited number of items during an online deployment stage, the performance cutting at a small value of $K$ is more important~\cite{DBLP:journals/tois/0007D0F0023}. 

\begin{figure}[t]
  \centering 
  \hspace{-0.4cm}
  \subfloat[]
  {
      \label{fig:expr1_1}\includegraphics[width=0.23\textwidth]{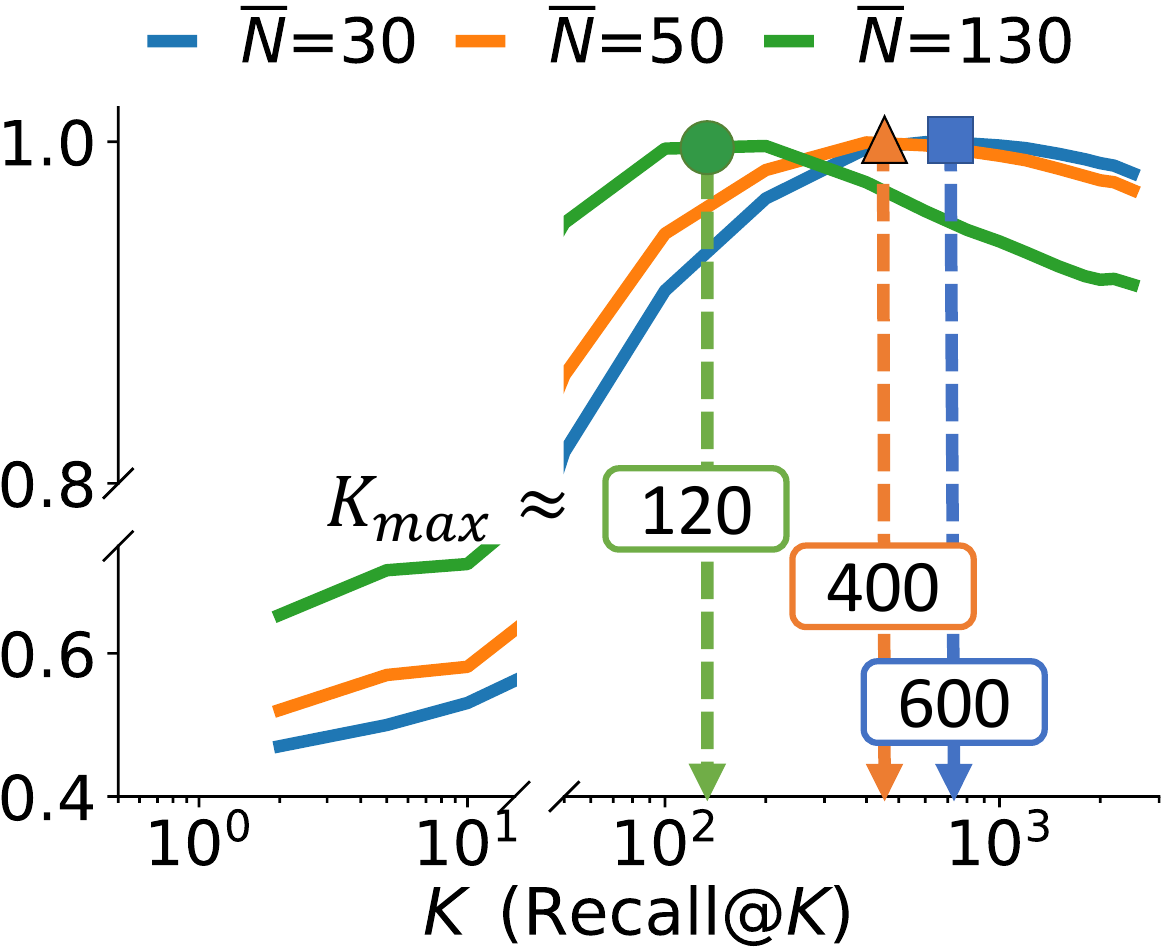}
  }
  \subfloat[]
  {
      \label{fig:expr1_2}\includegraphics[width=0.23\textwidth]{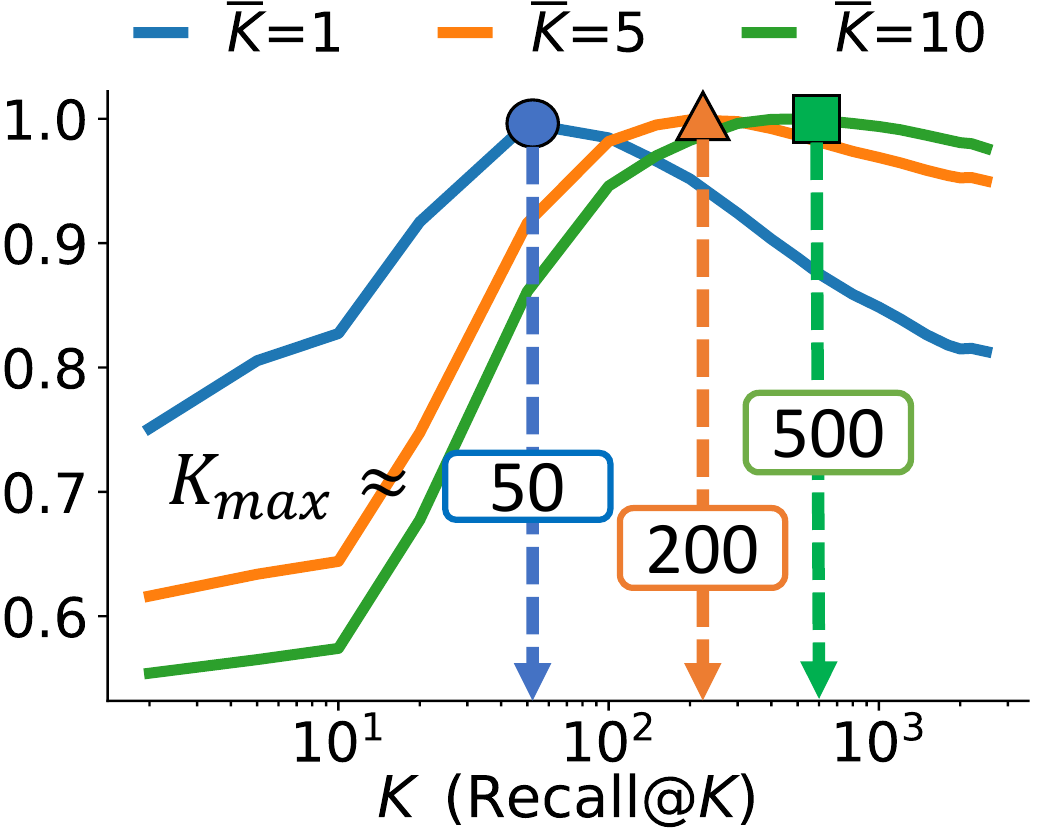}
  }
  \caption{The correlation coefficients between Recall@$\overline{K}$ on $D_{rand}$ and Recall@$K$ on $D_{full}$. 
  (a) The effect of $\overline{N}$, where $\overline{K}$ is fixed to 5. (b) The effect of $\overline{K}$, where $\overline{N}$ is fixed to 80. 
  }
  \label{fig:ranking}
\end{figure}

\subsection{Empirical Results}
\label{subsection:analysis expr}
In this subsection, we conduct empirical experiments to validate the relationship between Recall@$K$ on $D_{full}$ and Recall@$\overline{K}$ on $D_{rand}$.

We train multiple AutoDebias~\cite{DBLP:conf/sigir/ChenDQ0XCLY21} models w.r.t. different hyperparameters and evaluate their Recall@$\overline{K}$ on $D_{rand}$ and Recall@$K$ on $D_{full}$.  Then we calculate correlation coefficients between Recall@$\overline{K}$ and Recall@$K$ to measure their correlation. The results are shown in Figure~\ref{fig:ranking}. We highlight the value of $K$ when each curve reaches its maximum correlation coefficient as $K_{max}$. From the Figure~\ref{fig:ranking}, we have the following observations: 

\begin{itemize}[leftmargin=*, itemsep=2pt, topsep=2pt, parsep=2pt]
\item \textbf{When $\overline{N}$ and $\overline{K}$ are hold constant, the highest correlation coefficients between Recall@$\overline{K}$ and Recall@$K$ are achieved with large values of $K$.} As depicted in Figure~\ref{fig:expr1_1} and Figure~\ref{fig:expr1_2}, the Pearson correlation coefficient between Recall@$\overline{K}$ and Recall@$K$ surpasses 0.9 for values of $K$ exceeding 100, while it remains below 0.6 for values of $K$ below 10. This is consistent with our observation from Theorem~\ref{theorem_K} that the Recall@$\overline{K}$ on $D_{rand}$ only exhibits a strong correlation with the Recall@$K$ with large values of $K$ on $D_{full}$. Hence, the evaluation of models' Recall@$\overline{K}$ performance on $D_{rand}$ may not adequately reflect Recall@$K$ performance with small $K$ on $D_{full}$.
    
\item \textbf{When fixing $\overline{K}$, the value of $K_{max}$ decreases as the sample number $\overline{N}$ increases.} As shown in Figure~\ref{fig:expr1_1}, the inverse relationship between $K_{max}$ and $\overline{N}$ is consistent with the conclusion in Eq.~\eqref{eq:relation_Recall_K} that $K=\frac{N}{\overline{N}}\cdot \overline{K}$. This is intuitively understandable since an increase in the sample size $\overline{N}$ of $D_{rand}$ leads to a stronger and more reliable representation of the fully-exposed dataset $D_{full}$. However, in practice, scaling up the sample size is often impractical due to substantial economic constraints.

\item \textbf{When fixing $\overline{N}$, the value of $K_{max}$ decreases as the $\overline{K}$ decreases.} As shown in Figure~\ref{fig:expr1_2}, the positive correlation between $K_{max}$ and $\overline{K}$ is also consistent with the conclusion in Eq.~\eqref{eq:relation_Recall_K} that $\overline{K}=\frac{\overline{N}}{N}\cdot K$. However, as we discussed before, due to the frequent occurrence of $\overline{N}\ll N$, even when considering Recall@$\overline{K}$ with $\overline{K}=1$, it is still not possible to unbiasedly and accurately estimate Recall@$K$ for small values of $K$.
\end{itemize}
In conclusion, 
Recall@$\overline{K}$ is associated with Recall@$K$ when the value of $K$ is relatively large. 
Given that the gold standard for debiasing recommendation desires Recall@$K$ with small values of $K$ (\eg $K=50$), Recall@$\overline{K}$ is problematic.

\section{URE scheme}

In this section, we introduce a pioneering evaluation scheme coined URE, delineated in Figure~\ref{fig:new_method}. URE relies solely on $D_{rand}$ to achieve an unbiased estimation of Recall@$K$ on $D_{full}$, which consists of the following steps:
\begin{itemize}[leftmargin=*, itemsep=2pt, topsep=2pt, parsep=2pt]
    \item Obtain the predictions $\hat{y}$  ($\hat{y} = M (uid,iid)$) for all candidate items ($iid$$\in$[1, ..., N]), where the $M$ is the model to be tested.
    \item Sort these candidate items ($iid$$\in$[1, ..., N]) in descending order by their predictions $\hat{y}$, denoted as Sorted ItemID in Figure~\ref{fig:new_method}.
    \item Assign the labels (i.e. feedback) in $D_{rand}$  (positive : 1, negative : 0, Missing labels : \textbackslash)  to each candidate item in Sorted ItemID, denoted as Sorted $D_{rand}$ in Figure~\ref{fig:new_method}.
    \item Locate the ($K$+1)-th item in Sorted ItemID and its corresponding prediction $\hat{y}$.
    \item  Compute the $ \widehat{\text{Recall@}K}$ for this user: $ \widehat{\text{Recall@}K} = \frac{m}{n},$
    where $n$ denotes the number of items with positive label in Sorted $D_{rand}$, and $m$ denotes the number of these items with higher $\hat{y}$ than the ($K$+1)-th item.
    \item We report the averaged $\widehat{\text{Recall@}K}$ across all users as an unbiased estimate of the model's overall Recall@$K$ performance on $D_{full}$.
\end{itemize}

\noindent \textbf{Theoretical Guarantee}. The following theorem demonstrates that the estimates provided by URE are unbiased.



\begin{thm}
(URE Unbiasness) For all given $K$, we have:
    \begin{equation}
    E_{\mathcal{S}_{\overline{N}}} \left[\widehat{\text{Recall@}K} \right] = \text{Recall}@K,
    \end{equation}
where $\mathcal{S}_{\overline{N}}$ denotes the set of all possible random samplings of size $\overline{N}$.
\label{thm:unbias}
\end{thm}
\noindent This signifies that the estimates derived from URE are unbiased. 
\label{proof:unbias}
\begin{proof}
Since Recall is only related to positive samples, we can only consider positive samples. Let's assume that in $D_{full}$, there are a total of $N$ positive samples, out of which $M$ samples are ranked higher than the sample ranked at $K+1$ in Sorted ItemID in Figure~\ref{fig:new_method}. 
Assuming that we have randomly sampled $\overline{N}$ samples, which have $n$ positive samples and the rest are negatives. Since the sampling process is random, the probability of the number $m$ of positive samples that rank higher than the sample ranked at $K+1$ in Sorted ItemID can be calculated as $(C^{m}_{M}C^{n-m}_{N-M}/C^{n}_{N})$, thus we have:

\begin{figure}[tbp]
\setlength{\belowcaptionskip}{0cm}
  \centering
  \includegraphics[width=\linewidth]{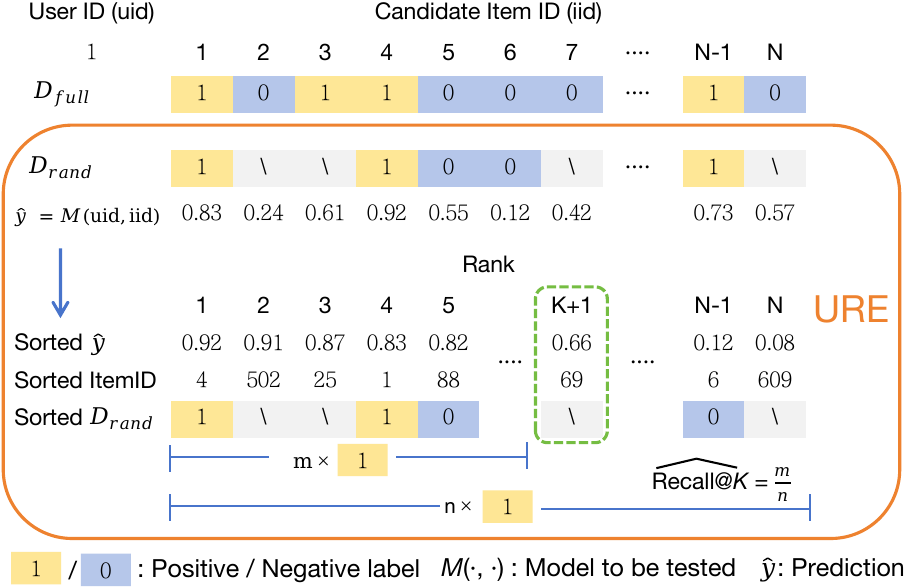}
  \caption{Illustration of the proposed URE scheme on unbiasedly estimating Recall@$K$ on $D_{full}$.}
  \label{fig:new_method}
\end{figure}

\begin{small}
\begin{align}
    E_{\mathcal{S}_{\overline{N}}} \left[\widehat{\text{Recall}@K} \right] 
    &=E_{n^{\prime}}\left[E_{\mathcal{S}_{\overline{N}}}[m/n | n^{\prime}] \right] \\
    & = E_{n^{\prime}} \left[\sum_{{m=1}}^{min(M, n^{\prime})} \frac{m n^\prime M}{ m n^{\prime} N} \frac{C^{m - 1 }_{M - 1} C^{n^{\prime} - 1 - (m - 1 )}_{N-M}}{C^{n^{\prime} - 1}_{N - 1}}\right] \\
    & = E_{n^{\prime}} \left[ \frac{M}{N} \sum_{{m=1}}^{min(M, n^{\prime})} \frac{C^{m - 1 }_{M - 1} C^{n^{\prime} - 1 - (m - 1 )}_{N-M}}{C^{n^{\prime} - 1}_{N - 1}}\right] \\
    & = M/N = \text{Recall}@K.
\end{align}
\end{small}
\end{proof}


\begin{figure}[tbp]   
  \centering 
  \hspace{-0.5cm}
  \subfloat[]{\label{fig:expr3_1}\includegraphics[width=0.24\textwidth]{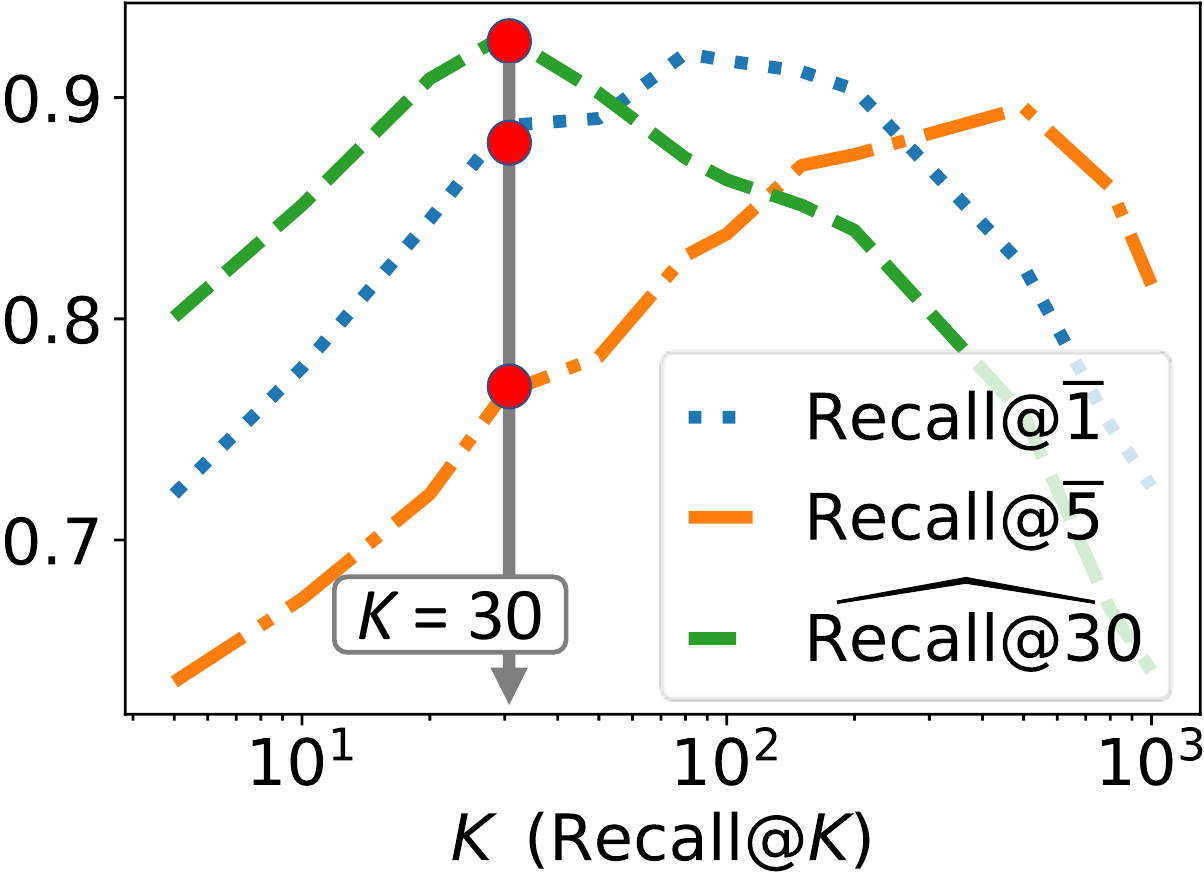}}	
  \hspace{0.2cm}
  \subfloat[]{\label{fig:expr3_2}\includegraphics[width=0.24\textwidth]{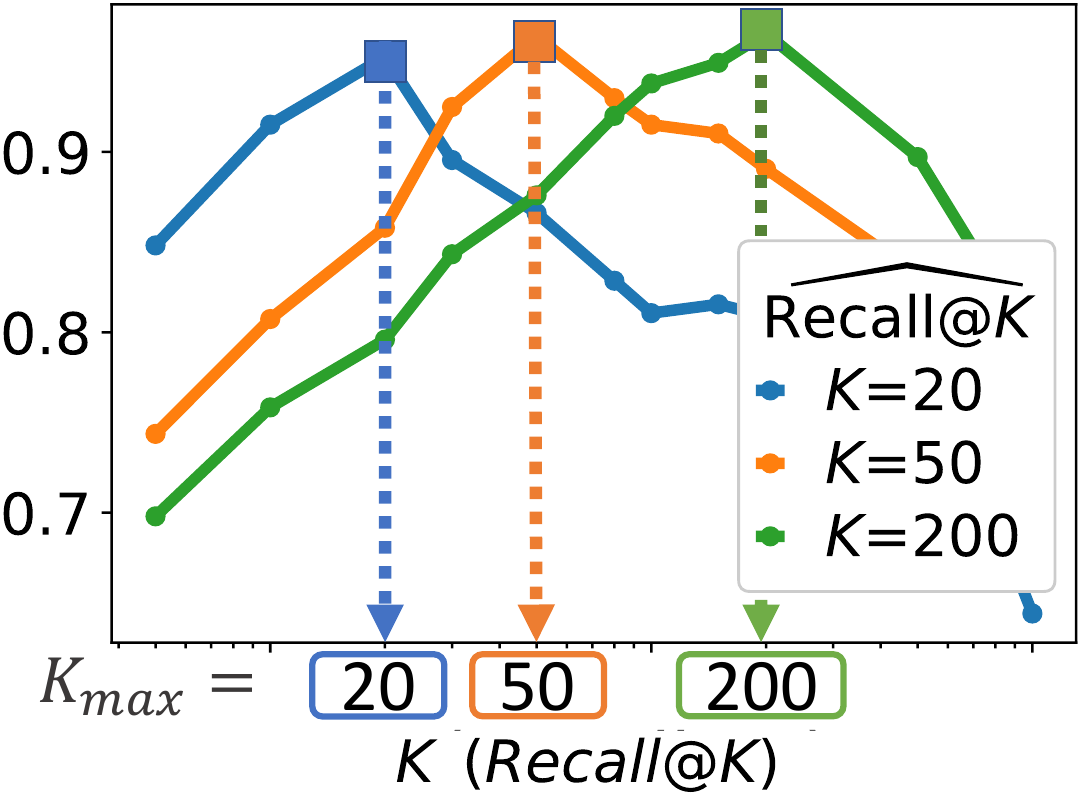}}
 
  \caption{(a) The correlation coefficients between Recall@$K$ and $\widehat{\text{Recall@}K}$ as well as the conventional Recall@$\overline{K}$. (b) The correlation coefficients between Recall@$K$ and $\widehat{\text{Recall@}K}$ for different values of $K$. 
  }
  \label{fig:expr_3}
\end{figure}
\section{Experiments}

\label{subsection:A}

\noindent \textbf{Unbiasedness Analysis.}
In this subsection, we test the AutoDebias model using the KuaiRec dataset to verify that the $\widehat{\text{Recall@}K}$ in the URE scheme is an unbiased estimation of Recall@$K$ on $D_{full}$.


In Figure~\ref{fig:expr3_1}, 
we calculate the correlation coefficient between Recall@$K$ and $\widehat{\text{Recall@}K}$, as well as Recall@$\overline{K}$ to demonstrate the ranking consistency. As shown in Figure~\ref{fig:expr3_1}, when $K$ is small (\textit{e.g.} $K=30$), both Recall@$\overline{1}$ and Recall@$\overline{5}$ exhibits poor correlations with Recall@$K$. In contrast, the correlation coefficient between the estimated $\widehat{\text{Recall@}K}$ and Recall@$K$ is higher than 0.9. This means that the model we choose based on $\widehat{\text{Recall@}K}$ is more likely to perform better on Recall@$K$ on $D_{full}$, which further demonstrates the effectiveness of the URE scheme.

In Figure~\ref{fig:expr3_2}, we report the correlation coefficient between Recall@$K$ and $\widehat{\text{Recall@}K}$ for different values of $K$. As we desired, only when K is the same, Recall@$K$ and $\widehat{\text{Recall@}K}$ have the highest correlation coefficients. That is, URE allows us to select the best debiased model based on the Recall@$K$ with a desired $K$.

Based on our discussion, we show that our URE scheme unbiasedly estimates Recall@$K$ for different values of $K$ and ensure an accurate debiased model selection.

\begin{table}[tbp] 
\centering
\renewcommand{\arraystretch}{1.2}
\caption{The evaluation results by URE. 
Notably, the Recall@$5$ is not available in Yahoo!R3, as it lacks a fully-exposed dataset. The results are statistically significant with $p <$ 0.01.}
\label{new_scheme_result}
\setlength{\tabcolsep}{1.mm}{

\resizebox{1\linewidth}{!}{
\begin{tabular}{cccc|cc}
\toprule
\multirow{3}{*}{Model}&\multicolumn{3}{c|}{KuaiRec} &\multicolumn{2}{c}{Yahoo!R3}\\
& \thead{\textbf{Traditional}\\ \textbf{scheme}} & \thead{\textbf{Fully-Exposed}\\ \textbf{Dataset}} &\thead{\textbf{URE}\\ \textbf{scheme}} & \thead{\textbf{Traditional}\\ \textbf{scheme}} &\thead{\textbf{URE}\\ \textbf{scheme}} \\
& Recall@$\overline{5}$ &  Recall@$5$  &$\widehat{\text{Recall@5}}$ &Recall@$\overline{5}$ &$\widehat{\text{Recall@5}}$\\ 
\cline{2-6}

MF              & 0.4859 & 0.0257 & 0.0267 & 0.7085 & 0.0131  \\ 
IPS             & 0.4889 & \underline{0.0266} & \underline{0.0273} & 0.6572 & 0.0095   \\ 
DR            & 0.4849 & 0.0262  & 0.0267 & 0.7195 & 0.0075  \\
AutoDebias    & \underline{0.5181}  & \textbf{0.0317} & \textbf{0.0323} & \underline{0.8085} & \textbf{0.0550} \\ 
DNS           & \textbf{0.5288} & 0.0237  & 0.0245 & \textbf{0.8330} & \underline{0.0490}  \\
\bottomrule
\end{tabular}
}
}
\vspace{-5pt}
\end{table}

\noindent \textbf{Re-evaluation Results.}
In this subsection, we re-evaluate some classical debiasing methods under backbone MF~\cite{MF}, including IPS~\cite{DBLP:conf/icml/SchnabelSSCJ16}, DR~\cite{DBLP:conf/icml/WangZ0Q19}, AutoDebias~\cite{DBLP:conf/sigir/ChenDQ0XCLY21}, DNS~\cite{DBLP:conf/www/ShiCFZWG023} on two widely-used datasets (\cf Table~\ref{tab:dataset}) in the debiasing research domain, and report the results in Table \ref{new_scheme_result}. 
We have the following key findings: (1) Based on the results on KuaiRec, we observe that the evaluation result $\widehat{\text{Recall@5}}$ of the URE scheme aligns closely with the true Recall@5 on $D_{full}$. This observation suggests a strong unbiasedness in the URE scheme. In contrast, the evaluation results of the traditional evaluation scheme exhibit a significant discrepancy from the true Recall@5, and it fails to ensure the preservation of order in the results. (2) When using $\widehat{\text{Recall@5}}$ to verify the debiasing performance, some debiasing methods perform worse than the orignal MF, which encourages the development of further debiasing techniques.




\begin{table}[tbp]
\setlength{\belowcaptionskip}{-0.2cm}
    \centering
    \caption{The statistics of datasets.}
    \label{tab:dataset}
    \small
    \resizebox{0.8\linewidth}{!}{
    \begin{tabular}{c c c c c}
        \hline
        Dataset &$\#$Users    &$\#$Items    &Biased train  &test   \\
        \midrule
        Yahoo!R3     &15,400 &1,000 &311,704 &48,600 ($D_{rand}$) \\
        KuaiRec     &1411 &3327 &1,934,404 &2,742,166 ($D_{full}$) \\
        \bottomrule
    \end{tabular}
}
\vspace{-5pt}
\end{table}

\section{Conclusion}

In this paper, we have unveiled the inconsistency between traditional evaluation scheme on the $D_{rand}$ and the gold standard evaluation on the $D_{full}$, which results in an inaccurate assessment of debiasing methods. To tackle this issue, we have introduced the URE scheme to unbiasedly estimate true Recall performance. Our proposal has been substantiated through both theoretical and empirical analyses, affirming its effectiveness.

\begin{acks}
This work is supported by the National Key Research and Development Program of China (2022YFB3104701) and the National Natural Science Foundation of China (62272437).
\end{acks}

\bibliographystyle{ACM-Reference-Format}
\balance
\bibliography{www}



\end{document}